\documentstyle[preprint,prb,aps,epsf]{revtex} %
\begin{document}

\draft

\title{
On the Bilayer Coupling in the Yttrium-Barium Family of High Temperature 
Superconductors
}

\author{\large A. J. Millis$^*$ $^a$ and H. Monien $^b$} \address{
$^a$
Bell Laboratories\\
Lucent Technologies\\
700 Mountain Avenue\\
Murray Hill, NJ 07974\\
$^b$
Physikalisches Institut\\
Universit\"at Bonn\\
Nu\ss allee 12\\
D-53115 Bonn\\
Germany
}

\maketitle

\begin{abstract}
We derive the expressions needed to interpret experiments relating to
interplane magnetic coupling in YBa$_{2}$Cu$_{3}$O$_{6+x}$ and related
materials, and use the results to interpret measurements of the optical magnon
energy in YBa$_{2}$Cu$_{3}$O$_{6.2}$ and of the NMR ``cross-relaxation" rate
in Y$_2$Ba$_{4}$Cu$_{7}$O$_{15}$. We estimate $J_{\perp}\sim 14$~meV in
both materials, and $\chi_{\rm max}/\mu^2_{B} \sim 100$ {\rm states}/{\rm
eV}-Cu in Y$_2$Ba$_4$Cu$_7$O$_{15}$ at $T=100$K. 
We show that there is at present no obvious contradiction between the
results of a widely-used analysis of NMR experiments and the results
of neutron scattering experiments in the Y-Ba system.  

We argue that the 41~meV excitation observed in superconducting
YBa$_{2}$Cu$_{3}$O$_{7}$ is a collective mode pulled down below the
superconducting gap by interactions, and that the observed
antisymmetry under interchange of planes follows from the
non-negligible value of $J_{\perp}$.

\end{abstract}

\vfill\eject

\section{Introduction}

In the yttrium-barium (Y-Ba) family of high temperature
superconductors the basic structural unit is a ``bilayer" consisting
of two CuO$_2$ planes; the bilayers are separated by CuO chains and
the coupling between bilayers is very weak. Neutron scattering
\cite{Tranquada92} and more recently NMR experiments \cite{Stern95}
have shown that the Cu spins on adjacent planes in a bilayer are
coupled. Intra-bilayer coupling has been shown theoretically to lead
to a ``spin gap"
\cite{Altshuler92,Millis93,Altshuler94,Ubbens94,Altshuler96} similar
to that observed \cite{Takigawa91} in NMR experiments on underdoped
members of the Y-Ba family. In view of the great importance of the
spin gap phenomenon, a quantitative analysis of the spin dynamics of a
bilayer is desirable. In this communication we provide this analysis
and use it to interpret NMR and neutron scattering experiments. Our
new results include 
(a) a calculation of the quantum renormalization of the optic
spin-wave gap in the $S={1\over 2}$ Heisenberg bilayer,
(b) the proper analysis of the NMR ``cross-relaxation" experiment
\cite{Stern95}, 
(c) refined estimates of the inter-bilayer coupling and in-plane
antiferromagnetic susceptibility,
(d) a demonstration that there is no obvious contradiction between the
results obtained from an analysis of the NMR data in terms of the
``MMP'' model \cite{Millis90,Barzykin95} and the results obtained from
neutron scattering \cite{Fong96} and 
(e) an improved understanding of a theory for the 41~meV peak observed
\cite{Mook93} in neutron scattering experiments on superconducting
YBa$_{2}$Cu$_{3}$O$_{7}$. A brief account of this work has been
presented elsewhere \cite{Millis95}

Focus on the two planes of a bilayer, and neglect coupling to other 
bilayers. Label the spin degrees of freedom by an index $a$ = 1, 2 
distinguishing planes, and a site index $i$. Consider the 
susceptibility
\begin{equation}
\chi^{ab}(q,\omega) =
\int_{0}^{\infty}dt\;
e^{i\left(\omega t + \vec q\cdot\left(\vec{R}_i-\vec{R}_j\right)\right)} 
\left<\left[\vec{S}_i^a(t), \vec{S}_j^b(0)\right]\right>. 
\end{equation}
Here $\vec{S}$ is the usual spin ${1\over 2}$ operator. Because of the 
symmetry under exchange of planes, $\chi^{ab}$ has only 
two independent components, $\chi^{11}(q,\omega)=\chi^{22}(q,\omega)$ and 
$\chi^{12}(q,\omega)=\chi^{21}(q,\omega)$. The two independent components 
of $\chi$ may be taken to be the even and odd (under the interchange of 
planes) components $\chi_{even, odd}=\chi_{11}\pm\chi_{12}$.
The main goals of this paper are to relate
experimental observations to $\chi^{11}$ and $\chi^{12}$ and these to
Hamiltonian parameters.

The rest of this paper is organized as follows. In section II we
derive an expression for the optic spin wave gap for the bilayer
Heisenberg model and use the result to obtain from data
\cite{Reznik96,Hayden96} 
an expression for the bilayer coupling $J_{\perp}$ in
YBa$_2$Cu$_3$O$_{6.2}$. In section III we formulate, solve and apply
to data a model for the spin dynamics of a metallic (but
non-superconducting) bilayer, obtaining estimates for $J_{\perp}$, the
antiferromagnetic susceptibility, and the temperature dependence of
the correlation length.  In section IV we extend the model to the
superconducting state and show that it accounts naturally for the
observation of the superconducting state 41~meV peak observed in
YBa$_{2}$Cu$_{3}$O$_{7}$ \cite{Mook93} and the lack of measurable
signal in the normal state. Our results provide a new interpretation
of a previously published numerical calculation
\cite{Liu95}, explain why the effect was not seen in another calculation
\cite{Normand95} and show that an objection raised 
by other workers 
\cite{Mazin95} is not 
valid. Section V is a conclusion.

\section{Bilayer Heisenberg Model}

In this section we estimate the optic magnon energy of the bilayer 
Heisenberg model defined by the Hamiltonian
\begin{equation}
H={1\over 2}J\sum_{{i\delta}\atop a}\, 
\vec{S}_{i}^{(a)}\cdot\vec{S}_{i+\delta}^{(a)}\, + \, 
J_{\perp}\sum_{i}\vec{S}_{i}^{1}\vec{S}_{i}^{2}
\label{eq:Heisenberg}
\end{equation}
Here $a=1,2$ labels planes, $i$ labels sites of a two dimensional square 
lattice, $\delta$ labels the four in-plane nearest neighbors of a site, 
and the $\vec{S}_{i}$ are the usual $S={1\over 2}$ spin operators.

We solve Eq. \ref{eq:Heisenberg} via the Schwinger boson method 
\cite{Arovas88}, which has been shown \cite{Singh89} to give very 
accurate estimates for the quantum renormalization of the single-plane 
$S={1\over 2}$ Heisenberg model. We previously used the Schwinger boson 
method to analyze the quantum disordered phases of Eq. 
\ref{eq:Heisenberg} \cite{Millis94}. The analysis presented here amounts 
to an extension of that work to the ordered phase; 
we therefore omit all details of the derivation of the equations, and 
mention only necessary notation and new features.

In brief, to use the Schwinger boson method one introduces Bose fields 
$b^{+ (a)}_{i \alpha}$ so that $S^{(a)}_{i \alpha} = b_{i \alpha}^{\dagger(a)} 
\vec{\sigma}_{\alpha\beta}b_{i \beta}^{\phantom{\dagger}(a)}$.
The fields are subject to the constraint 
$(2S+1)=
\sum_{\alpha}b_{i\alpha}^{\dagger(a)} b_{i \alpha}^{\phantom{\dagger}a}$. 
The $\vec{S}\cdot \vec{S}$ interactions become four-boson interactions,
which one decouples with Hubbard-Stratonovich fields $Q$ (for the
in-plane interaction $J$) and $\Delta$ (for the between planes
interaction $J_{\perp}$). One must also introduce a Lagrange
multiplier $\mu$ to enforce the constraint.  The mean field equations
are given in Eqs. 2.9 of Ref.
\cite{Millis94} and are rewritten here for convenience: 
\begin{eqnarray}
\int{d^{2}k\over (2\pi)^2}{\mu\over \omega_{k}}
{\left[1+2b({\omega_{k}\over T})\right]}
& = 1+2S \nonumber \\
\int{d^{2}k\over (2\pi)^2}
{(Q\gamma_{k}+\Delta)\gamma_{k}\over \omega_{k}}
{\left[1+2b({\omega_{k}\over T})\right]}
& = 
Q/2J \nonumber \\
\int{d^{2}k\over (2\pi)^2} 
{(Q\gamma_{k}+\Delta)\gamma_{k}\over \omega_{k}}
{\left[1+2b({\omega_{k}\over T})\right]}
& = 2\Delta/J_{\perp} 
\label{eq:MF}
\end{eqnarray}

Here $\gamma_{k} = -{1\over 2}(\cos k_{x}+\cos k_{y}), \omega_{k} = 
\sqrt{\mu^{2} - (Q\gamma_{k}+\Delta)^2}$ and $b(x) = (e^{x}-1)^{-1}$. In 
the limit $J_{\perp}\rightarrow 0$ these equations reduce those studied 
by Arovas and Auerbach \cite{Arovas88}, except
that in our conventions  $Q$ 
is four times larger than in theirs. At low $T$
for parameters such that the ground state is ordered at $T=0$
one has
\begin{equation}
\mu^{2}=(Q+\Delta)^{2}+(\kappa/2)^{2}
\label{eq:mu}
\end{equation}
with inverse correlation length $\kappa$ given by 
\begin{equation}
\kappa = \kappa_{0} \exp( - 2\pi \rho_{s}/T).
\label{eq:kappa}
\end{equation}
Equation \ref{eq:kappa} defines the spin stiffness $\rho_{s}$.

In the $T\rightarrow 0$ limit the integrals in Eqs. \ref{eq:MF} may be 
evaluated up to corrections of order $T$ by noting that the terms 
proportional to $b(\omega_{k}/T)$ are dominated by the divergence as 
$\omega_{k}\rightarrow 0$ while in the terms proportional to 1 one may set 
$\kappa=0$. If $T$ is less than the optic mode gap one finds
\begin{eqnarray}
Q &= &2J[2S+1-I_{oa}+I_{ob}]\nonumber \\
\Delta &= &{1\over 2}J_{\perp}[2S+1-I_{oa}+I_{oc}]\nonumber \\
\rho_{s} &= &{1\over 4}Q[2S+1 - I_{oa}]
\label{eq:MFresults}
\end{eqnarray}

Here $I_{o a,b,c}$ are respectively
the $T=0$ values of the first, second and third 
integrals on the left hand 
sides of Eqs. \ref{eq:MF}. They depend upon the ratio $\Delta/Q$ and 
in the 
limit $\Delta \ll Q$ are 
\begin{eqnarray}
I_{oa} &=& 1.39 -{2\over\pi}\sqrt{\Delta\over Q} + O\left({\Delta\over 
Q}\right)\nonumber \\
I_{ob} &=& 0.55 -{2\over\pi}\sqrt{\Delta\over Q} + O\left({\Delta\over 
Q}\right)\nonumber \\
I_{oc} &=& {2\over\pi}\sqrt{\Delta\over Q} + O\left({\Delta\over Q}\right)
\label{eq:I}
\end{eqnarray}

Expressions for dynamic susceptibilities are given in Appendix C of Ref.
\cite{Millis94}. In the limit of interest here the absorptive parts are 
dominated by poles representing undamped spin waves.  In the acoustic 
sector the dispersion is given by $\omega=\omega_{k}$ with 
$k$ near 
$Q=(\pi ,\pi)$. 
In the optic sector the dispersion for wavevectors near $Q$ is given by 
$\omega=\omega_{k+Q}$.
In particular the acoustic spin wave velocity is $c^{2}={1\over 
2}Q(Q+\Delta)$ and the optic-mode gap $\omega_{\rm opt}=2\sqrt{\Delta Q}$. 
Neglecting terms of relative order $\Delta /Q$ we find 
$c=2\sqrt{2}Z_{c}(S)JS$ and $\rho_{s}=2Z_{\rho_{s}}(S)JS^{2}$ with 
$Z_{c}({1\over 2})=1.16$ and $Z_{\rho_{s}}({1\over 2})=0.71$, 
in agreement 
with previous work \cite{Arovas88,Singh89}. 
[The factor of two in $\rho_{s}$ occurs because each 
unit cell contains two $Cu$ ions, one in each plane; for $T \ll 
\omega_{opt}$ the two planes move together].

The equation for $\Delta$ may be rearranged to read
\begin{equation}
J_{\perp}={\omega_{opt}^2\over 8\rho_{s}+{2\over\pi}\omega_{opt}}
\label{eq:Jperp}
\end{equation}

If one takes the limit $J_{\perp}/\rho_{s}\rightarrow 0$ and sets 
$Z_{\rho_{s}}=1$,
this reduces to the usual $\omega_{opt}=4\sqrt{J J_{\perp}}S$.

Recent neutron scattering measurements \cite{Reznik96,Hayden96} have 
reported $\omega_{opt} \cong 70\, {\rm 
meV}$
for YBa$_{2}$Cu$_{3}$O$_{6.2}$. 
Use of the canonical values $J=100\, {\rm 
meV}$ 
and $Z_{\rho_{s}}=0.7$ implies $J_{\perp}\cong 14\, {\rm meV}$.  
Our value for 
$J_{\perp}$ differs from that quoted in the experimental
papers because those works assumed 
$Z_{\rho_{s}}=1$ and neglected the second term in the denominator of Eq. 
\ref{eq:Jperp}.  Some caution is necessary in using this value because 
the measurements were made on samples of nominal composition 
YBa$_{2}$Cu$_{3}$O$_{6.2}$. The effects of the extra $0.2$ of oxygen might 
lead to substantial changes of $Z_{\rho_{s}}$ from the ideal-system
$Z_{\rho_{s}}\approx 0.7$ value.

We can also obtain from the formalism an expression
for the strength of the spin-wave poles.
In the regime of interest the momentum 
integrals are dominated by the infrared divergence at $\omega_{k}\ll T$;
we find, for the susceptibility per plane,
\begin{equation}
\chi^{''}(q,\omega)={16\pi \rho_{s}\over 
\omega_{q}}[\delta(\omega-\omega_{q})-\delta(\omega+\omega_{q})]
\label{eq:chispinwave}
\end{equation}

For the acoustic branch, $\omega_{q}=cq$; for the optic branch, 
$\omega_{q}=\omega_{opt}$. The prefactor may be written as $4\pi S\sqrt{2} 
Z_{\rho_{s}}/Z_{c}(qa)$ or (using $c^{2} = \rho_{s}/\chi_{\perp}$) 
$4\pi S\sqrt{2} Z_{\chi}Z_{c}/(qa)$, where $a$ is the lattice constant. 
This susceptibility is, up to an apparently omitted factor of $Z_c$, 
consistent with that given in Eq. 8 of \cite{Fong96}. Those authors define 
$\chi$ as ${1\over2\pi}\langle S^{-}S^{+}\rangle$ and measure it per 
bilayer. The factor of two from $S^{-}S^{+}$ versus $S^{z}S^{z}$ and the 
factor of two from bilayers leads to
a difference of $\pi/2$ in our conventions.

\section{Metallic Bilayer}

This situation is more complicated and less well defined than the 
insulating
antiferromagnetic case because there is neither a generally accepted 
theory of the magnetic dynamics of a single layer nor a generally 
accepted theory of the interlayer coupling. One must therefore proceed
phenomenologically. Despite the obvious limitations, we believe that this 
is worth doing because the spin dynamics of metallic cuprates are a 
subject of continuing interest and, as we shall show, recent measurements 
of interplane effects yield information about both the between-planes 
coupling and the in-plane spin dynamics.

Further analysis requires a model. We shall assume that the magnetic 
dynamics of a single plane may be described by a susceptibility 
$\chi_{0}(q,\omega)$ whose form we discuss further below. We also assume 
that the only interplane coupling is the magnetic one, 
$J_{\perp}{\displaystyle \sum_{i}}\vec{S}_{i}^{(1)}\cdot\vec{S}_{i} 
^{(2)}$.  This is not an important restriction.  Retaining
the off-diagonal term $\chi_0^{12}$ in our formalism
leads only to a modest renormalization of $J_{\perp}$,
as shown also in \cite{Liu95}. Finally, we assume that
the effects of $J_{\perp}$ may be modeled via the RPA. Thus we write
\begin{equation}
  \bbox{\chi}^{-1} =\left[ 
    \begin{array}{cc}
      \chi^{-1}_{0}(q,\omega) & -J_\perp \\
      -J_\perp & \chi^{-1}_{0}(q,\omega)
    \end{array}
  \right].
  \label{eq:chi-rpa}
\end{equation}
We further assume (as has been done in previous analyses of NMR in
high ${\rm T_c}$ materials \cite{Millis90}) that $\chi_{0}(q,\omega)$
has the scaling form $\chi_{0}(q,\omega) = \chi_0 \xi^2
f(q\xi,\omega/\xi^z)$, where $f$ is a scaling function normalized so
that $f(0,0)=1$, $z$ is the dynamical exponent, $\xi$ is a correlation
length measured in units of the lattice constant and $\vec q$ is
measured from an ordering wavevector $\vec Q$ which for the present
discussion is arbitrary. $\vec Q$ is believed to be of the order of
$(\pi,\pi)$ in high T$_{\rm c}$ materials.  {}From Eq.
\ref{eq:chi-rpa} we see that $\chi^{11}(q,\omega) =
\chi_0(q,\omega)/(1-(J_\perp\chi_0(q,\omega))^2)$ and
$\chi^{12}(q,\omega) = J_\perp
\chi_0(q,\omega)^2/(1-(J_\perp\chi_0(q,\omega))^2)$. The crucial
parameter controlling the susceptibilities in the static limit is
\begin{equation}
I_\perp \equiv J_\perp \chi_{max} =J_\perp\chi_0\xi^2. 
\label{eq:Delta} \end{equation}
Here $\chi_{max}$ is the maximum value
of the in-plane spin susceptibility. We must assume 
$I_\perp<1$ so that the material has no long range order. 
If $I_\perp^2 \ll 1$ then $\chi^{11} \approx \chi_0$ and $\chi^{12} 
\approx J_\perp \chi_0^2$. In this limit the interplane coupling has a 
weak effect and the RPA (which in this limit
is just a perturbation expansion in $J_{\perp}$)
 is an appropriate model. On the other hand, if 
$I_\perp ^2 \approx 1$ then the interplane coupling is strong and the use 
of the RPA may be questioned. 

We now turn to the NMR experiments of interest. These are $T_2$ 
experiments performed on Y$_2$Ba$_4$Cu$_7$O$_{15}$, a material in which 
the single-chain structure of YBa$_2$Cu$_3$O$_7$ alternates with the 
double-chain structure of YBa$_2$Cu$_4$O$_8$ \cite{Stern95,Stern94}. 
As a result, 
atoms on different planes of a bilayer have somewhat different local 
environments and therefore somewhat different NMR resonance frequencies, 
which may be independently studied. Despite the differences in local 
environment, the electronic properties of the two planes seem essentially 
identical -- in particular, ratios of Knight shifts and in-plane 
relaxation rates are temperature-independent \cite{Stern94}.
We therefore believe that the differences in observed
relaxation rates and Knight shifts are due to differences
in hyperfine couplings.  An alternative view is that
the electronic susceptibilities on the two planes differ.  This
difference may be included in our formalism and analysis, and
will not change the results in an important way (basically, one
replaces $\chi^{11}$ by $(\chi^{11}\chi^{22})^{1/2}$).

Now the NMR $T_2$ measures the rate 
at which a nuclear spin is depolarized by interacting with other nuclear 
spins, i.e. it measures the nuclear-spin--nuclear-spin interaction 
strength. In high $T_c$ materials the dominant contribution to the 
nuclear-spin nuclear-spin interaction comes from polarization of 
electronic spins, and may be related to the static limit of the real part 
of the electronic spin susceptibility
\cite{Pennington91}. In Y$_2$Ba$_4$Cu$_7$O$_{15}$ it is 
possible to measure $T_{2}$, the rate at which a spin in one plane is 
depolarized by spins in the same plane, and $T_{2\perp}$, the rate at 
which a spin in one plane is depolarized by spins in the other plane. 
$T_{2}$ is related to the electronic spin susceptibility by 
\cite{Thelen94}
\begin{equation}
\frac{1}{T_{2}^{(a)}}=
\left[
\sum_q\left[{A_q^{(a)}}^2 \chi^{11}_q\right]^2-
\left(\sum_q {A_q^{(a)}}^2 \chi^{11}_q\right)^2
\right]^{1/2}
\label{eq:T2}
\end{equation}
while $T_{2\perp}$ is given by
\cite{Monien94}
\begin{equation}
\frac{1}{T_{2\perp}} =
\left[\sum_q\left[A_q^{(1)}A_q^{(2)} \chi^{12}\right]^2\right]^{1/2}. 
\label{eq:T2perp}
\end{equation}

Here we have allowed for the different hyperfine couplings in the two
planes observed experimentally. We have calculated $T_{2}$ and
$T_{2\perp}$ from
Eqs. (\ref{eq:chi-rpa},\ref{eq:T2},\ref{eq:T2perp}). The precise
values obtained depend upon the form chosen for $f(q\xi,\omega=0)$. We have
used two forms for $f(x) \equiv f(q\xi,\omega=0)$: a Lorentzian,
$f(x)=1/(1+x^2)$, and a Gaussian, $f(x)=\exp(-\log(2) x^2)$ (the
$\log(2)$ is introduced so $f(x=1)=1/2$).  We measure $\xi$ in units
of the lattice constant, we set $\hbar=1$ and assume that the
hyperfine couplings can be approximated by their values at $Q$,
$A_Q^{a}$. We find
\begin{equation}
\frac{1}{T_{2}^{(a)}} = {A_Q^{(a)}}^2 \chi_0 \xi g_{in}(I_\perp) 
\label{eq:T2in-ans} 
\end{equation}
and
\begin{equation}
\frac{1}{T_{2\perp}} = A_Q^{(1)} A_Q^{(2)}J_\perp \chi_0^2 \xi^3 
g_{\perp}(I_\perp) \label{eq:T2perp-ans}
\end{equation}
where
$g_{in}$ and $g_\perp$ are defined in terms of the function $f(x)$ via 
\begin{equation}
g^2_{in}(I_\perp) = \frac{1}{2\pi} \left[ \int_0^\infty dx\; 
\frac{f^2}{(1-I_\perp^2 f^2)^2} - \frac{1}{2\pi\xi^2} \left( \int_0^\infty 
dx\; \frac{f^2}{1-I_\perp^2 f^2} \right)^2 \right]
\label{eq:gin}
\end{equation}
\begin{equation}
g^2_\perp(I_\perp) = \frac{1}{2\pi}
\int_0^\infty dx\; \frac{f^4}{(1-I_\perp^2 f^2)^2}.
\label{eq:gperp}
 \end{equation} In writing Eqs. \ref{eq:gin},\ref{eq:gperp} we have
assumed that the correlation length is so long that lattice effects
may be neglected. We have investigated this issue by performing the
exact integrals numerically. The parameter governing the size of the
lattice effects is $(\pi \xi)^{-1}$; for $\xi
\ge 1$ we have found that lattice effects are negligible. 

In Fig. 1 we present the calculated results for
$\sqrt{T_2^{(1)}T_2^{(2)}}/T_{2\perp}$ $=I_\perp g_{\perp}(I_\perp)
/g_{in}(I_\perp)$. From the experimental values \cite{Stern95} 
$\sqrt{T_2^{(1)}T_2^{(2)}}/T_{2\perp} =0.15$ at $T$ = 200 K and
$\sqrt{T_2^{(1)}T_2^{(2)}}/T_{2\perp} =0.30$ at $T$ = 120 K
we obtain $I_\perp \approx 0.2$ at $T$ = 200 K and
$I_\perp \approx 0.4$ at $T=120$ K. Note that even at the lowest
temperature we find that the system is in the small $I_\perp$ regime in
which $T_2/T_{2\perp}$ is linear in $I_\perp$, suggesting that the
between-planes coupling is sufficiently small that the RPA formula is
justified.

The magnitude of $J_\perp$ may be determined if $\chi_{max}$ is known
and conversely.  In the NMR and bulk susceptibility literature $\chi$
is defined in terms of the $[g\mu_B\vec{S},g\mu_B\vec{S}]$ correlator,
and results are presented as $\chi/\mu_B^2$.  This quantity differs by
a factor of four from the susceptibility defined in this paper; we
refer to it as $\chi^{exp}$.  If the susceptibility were only weakly
$q$-dependent then the measured uniform susceptibility
$\chi^{exp}_{\rm uniform} \approx$ 2 states/eV-Cu \cite{Walstedt92}
would provide a good estimate for $4\chi_{max}$ and our value $I_\perp
\approx 0.4$ would imply $J_\perp \approx$ 0.8 eV. Such a value is very 
difficult to justify on microscopic grounds because the insulating
antiferromagnetic parent compounds of the high $T_c$ superconductors
have in-plane exchange constants $J_{\rm in-plane} \approx 0.12$ eV
and, as shown in the previous section, $J_{\perp}/J_{\rm in-plane}\sim
0.1$. It seems very unlikely that doping would lead to a $J_{\perp}\gg
J_{\rm in-plane}$. Therefore, we believe the cross-relaxation results
imply $\chi_{max} \gg
\chi_{uniform}$ in metallic, superconducting materials. 
A similar conclusion has been drawn from analyses of
in-plane NMR \cite{Millis90,Pennington91,Barzykin95}, but these
analyses required assumptions about magnitudes of hyperfine couplings
and correlation lengths. These quantities drop out of the present
analysis. The estimate $J_\perp \approx 10-20$~meV has been obtained
from band structure calculations \cite{Andersen94}, implying
$\chi^{exp}_{max} \approx$ 80-160 states/eV-Cu, while in
Ref. \cite{Stern94} the estimate $J_{\perp} \approx 25 meV$ was
presented.

We now turn to a more detailed discussion of the value of $\chi_{max}$
inferred from the discussion of the in-plane NMR. This discussion is
based on the results contained in the recent paper of Barzykin and
Pines \cite{Barzykin95} which summarizes and analyzes a wide range of
NMR data. (The results proposed in Ref. \cite{Zha96} lead to
corrections which are smaller than our estimated uncertainties so we
have not incorporated them).  This paper is extremely useful, but has
a few limitations: in particular, explicit error estimates are not
given, and the analysis for YBa$_2$Cu$_3$O$_{6.63}$ does not include
interplane effects argued below to be important. We believe the
uncertainties in values for YBa$_2$Cu$_3$O$_{6.63}$ are substantial
and difficult to estimate with accuracy.

Barzykin and Pines show that once the hyperfine coupling is
determined, the measured in-plane $T_{2}^{-1}$ rate is proportional to
the quantity $\chi_{max}/\xi$. Their Table I and Fig. 2 gives for
YBa$_2$Cu$_3$O$_7$ the estimate $\chi_{Q}^{exp}(T=100K)/\xi=$30
states/eV-Cu.  We believe 10\% uncertainties in hyperfine couplings
are reasonable \cite{Millis92}, implying 20\% uncertainties in
$\chi_{max}^{exp}/\xi$.  To obtain an estimate for $\chi_{max}$ an
estimate for $\xi$ is required. We argued \cite{Millis92} that the
uncertainties in the NMR estimate for $\xi$ are large; we suggested,
on the basis of earlier data, that for YBa$_2$Cu$_3$O$_7$ at T=100K a
plausible estimate is $0.5$ \raisebox{-.2ex}{$\stackrel{\textstyle
<}{\sim}$} $\xi$ \raisebox{-.2ex}{$\stackrel{\textstyle <}{\sim}$} $2$
with the most likely value $\sim 1-1.5$ \cite{Millis93}.  From Fig. 14
and Fig. 16 of Ref. \cite{Barzykin95}, which are based on more recent
data, one obtains the larger estimates $\xi /a = 1.9-2.3$.  For
YBa$_2$Cu$_3$O$_{6.6}$ it seems clear that $\xi$ must be larger than
in YBa$_2$Cu$_3$O$_7$; The analysis of Ref. \cite{Barzykin95} yields
$\xi/a(T=100K) = 7.4$.  This value seems to us too large, but
regardless of the precise value of $\xi$, it is clear that the larger
T$_2^{-1}$ and presumably larger $\xi$ implies that $\chi_{max}$ in
YBa$_2$Cu$_3$O$_{6.63}$ is larger by a factor at least of order four
than in YBa$_2$Cu$_3$O$_7$.We thus roughly estimate that
$\chi_{max}^{exp}$ in YBa$_2$Cu$_3$O$_7 \sim$ 50 states/eV$-$Cu (with
$\sim$ factor of two uncertainties) and $\chi_{max}^{exp}$ in
YBa$_2$Cu$_3$O$_{6.63}$ is at least 250 states/eV-Cu (with rather
larger uncertainties).  The value obtained from the results in
Ref. \cite{Barzykin95} is $\chi_{max}^{exp}\sim$ 440 states/eV-Cu.  As
previously remarked, Y$_2$Ba$_4$Cu$_7$O$_{15}$ has a doping
intermediate between the two materials, but probably closer to
O$_7$. The previously derived estimate $\chi_{max}^{exp} \sim 80-160$
states/eV-Cu is thus consistent with the results of the in-plane NMR,
so it is likely that the effective between-planes coupling does not
change much with doping.

The rapid dependence of $\chi_{max}$ with doping inferred from
in-plane NMR measurements suggest that for YBa$_2$Cu$_3$O$_{6.63}$ and
lower dopings, $\chi_{max}$ is so large that $I_\perp \approx 1$. For
these materials the planes are thus very strongly coupled at low
energies. The simple RPA formula we have used is unlikely to be
accurate, and the between-planes coupling must be taken into account
in interpreting even in-plane NMR experiments. A preliminary attempt
along these lines has appeared
\cite{Millis94}.  

In summary, the cross-relaxation experiment shows that the real part
of the susceptibility at some non-zero wavevector $q$ is much larger
than the uniform susceptibility.  The temperature dependence of the
$T_2$ rates must be due to the temperature dependence of this
antiferromagnetic maximum. Two scenarios have been proposed for the
temperature dependence: in the {\it antiferromagnetic scenario} the
temperature dependent quantity is the correlation length $\xi$. In the
{\it generalized marginal Fermi liquid scenario} the temperature
dependent quantity is the overall amplitude $\bar{\chi}\;$
\cite{Si94}. From Eqs.  (\ref{eq:T2in-ans},\ref{eq:T2perp-ans}) we see
that in the regime where $T_{2\perp}$ is linear in $I_\perp$ the
antiferromagnetic scenario predicts $T_{2}^3/T_{2\perp}$ is
temperature independent, while the marginal Fermi liquid scenario
predicts $T_{2}^2/T_{2\perp}$ is temperature independent. The
experimentally determined ratios are plotted in Fig. 2 and are more
consistent with the antiferromagnetic scenario.

The imaginary parts of the two independent susceptibilities
$\chi_{even}$ and $\chi_{odd}$ are measurable via neutron scattering
because they have different dependences on $q_z$, the momentum
transverse to the CuO$_2$ planes \cite{Tranquada92}. Neutron
scattering experiments have been performed on a variety of metallic
members of the yttrium-barium family of high-T$_{\rm c}$ materials
\cite{Tranquada92,Mook93,Rossad-Mignod91}. The experimental 
result is that {\it only} $\chi_{odd}$ is seen. At frequencies less
than 30~meV and temperatures less than room temperature the even
parity fluctuations are claimed to be completely frozen out.  The
theory of neutron scattering in high $T_c$ materials is presently
controversial. There is no generally accepted model which correctly
accounts for the observed lineshapes and temperature dependences. To
investigate the connection between the cross-relaxation experiments
and neutron scattering we have chosen to calculate the ratio of the
$q$-integrated even and odd parity susceptibilities. This ratio is
insensitive to the precise details of the susceptibilities. For
definiteness we used the ``MMP", dynamical exponent $z=2$ ansatz
$\chi_0(q,\omega) = {\bar \chi} /(\xi^{-2} + q^2 -
i\omega/\Gamma)$. Here $\Gamma$ is a microscopic spin relaxation
time. The results depend on $I_\perp$ and on $\omega_{SF} =
\Gamma/\xi^2$, which we define here to be the softest spin fluctuation 
frequency of a single plane. Of course the between-planes coupling
will reduce this frequency for the odd parity channel and increase it
for the even channel. Results are shown in Fig. 3 for several values
of $I_\perp$.  We see that the relative weight of the even parity
fluctuations becomes small only for $I_\perp > 0.5$. We believe that
the neutron results, which seem to require a $I_\perp > 0.5$, are not
in contradiction to our analysis of the cross-relaxation experiment,
which yielded a $I_\perp \le 0.4$, because the strongest neutron
evidence for locked bilayers was obtained from a study of
YBa$_2$Cu$_3$O$_{6.5}$ \cite{Tranquada92}, which as we have previously
noted is much closer to the magnetic instability than
Y$_2$Ba$_4$Cu$_7$O$_{15}$, and therefore may be expected to have 
$I_\perp \sim 1$.

As a side remark we note that a recent neutron scattering work
provided a bound on the magnetic scattering in YBa$_2$Cu$_3$O$_7$
\cite{Fong96}.  Bearing in mind the factor of ($g^2\pi/2$) difference
between NMR and neutron conventions discussed above, we note that the
upper bound implied by the neutron experiment for $YBa_2Cu_3O_7$ is
$\chi_{max}^{exp}<300$ states/eV-Cu.  This bound is comfortably larger
than the NMR estimates obtained by interpreting in-plane NMR
experiments. At this time, therefore, there is no evidence of a
contradiction between neutron scattering experiments and the ``MMP''
scenario for interpreting the NMR.

\section{Superconducting State}
 
Another experiment relevant to bilayer coupling in the Y-Ba system is the 
observation of a sharp peak at an energy of 41~meV in the antisymmetric 
susceptibility of {\it superconducting} YBa$_2$Cu$_3$O$_7$ \cite{Mook93}.
A corresponding peak in the {\it symmetric} channel has not been observed 
\cite{Mook93,Fong96}, and an upper bound on the intensity in the 
symmetric channel of about 30{\%} of the intensity in the antisymmetric 
channel has been established \cite{Fong96}. A number of theoretical works 
have investigated this peak (see, e.g. 
\cite{Liu95,Mazin95,Demler95,Kohno96} and references therein). 
We believe that the most likely explanation is that the peak is a
collective mode pulled down below the superconducting gap edge by
interactions, and that its appearance only in the antisymmetric
channel is due to the non-zero $J_{\perp}$.  This explanation is
essentially that given by Liu et al. \cite{Liu95} (although they did
not interpret their results -- which were basically numerical -- in
precisely this manner).  An objection to this interpretation was
raised by Mazin and Yakovenko \cite{Mazin95}, who argued that within
such theories an observable peak should also exist in the symmetric
channel.  In the remainder of this section we present a simple RPA
theory of the 41~meV mode. Although the RPA is presumably not
quantitatively accurate it is in fact the basis of most previous work
(\cite{Liu95,Normand95,Mazin95}) and is to some extent analytically
tractable. Our treatment gives insight into the results of
\cite{Liu95}, shows why the effect was not seen by \cite{Normand95}
and shows that the objection of \cite{Mazin95} is unfounded.

The fundamental object in the RPA is the bare electron polarizability
$\chi _{00}$. This is given, at $T=0$ in a superconductor, by
e.g. Eq. 1 of Ref. \cite{Mazin95} (although one must multiply their
expression by 2 to convert to our conventions).  In two spatial
dimensions the constraints that the initial and final states lie on
the Fermi surface completely specifies the kinematics (up to discrete
lattice symmetry operations). Thus the threshold behavior of
$\chi^{''}$ for frequencies near the gap edge may be found
analytically. For $Q=(\pi, \pi)$ and bearing in mind that $2Q$ is a
reciprocal lattice vector one finds
\begin{eqnarray}
\chi_{00}(Q,\omega) &=&\chi^{\rm reg}_{00}+
\frac{4}{\pi}
{
\frac
{\sqrt{\left|\Delta_{p_0}\Delta_{p_0+Q}\right|}} 
{v_{p_{0}}v_{p_{0+Q}} \sin\theta}
}
\log{
\left|
\frac
{\sqrt{\left|\Delta_{p_0}\Delta_{p_0+Q}\right|}}
{\omega-\Delta_{p_0}-\Delta_{p_0+Q}}
\right|
}
\nonumber \\
{}\nonumber \\
&\equiv &
\chi_{00}^{\rm reg}
\left[ 
1 + A 
\log{
\left|
\frac
{\sqrt{\left|\Delta_{p_0}\Delta_{p_0+Q}\right|}}
{\omega-\Delta_{p_0}-\Delta_{p_0+Q}}
\right|\,
} 
\right]
\label{eq:chinonint}
\end{eqnarray}

Here $p_{0}$ is a wavevector such that
$\epsilon_{p_0}=\epsilon_{p_0+Q}=0$, $\theta$ is the angle between
$\vec{v}_{p_0}$ and $\vec{v}_{p_0+Q}$ and it is assumed that the Fermi
surface does not pass through a van~Hove point (so $v_{p_0}\neq 0$) and
$\theta\neq 0, \pi$.  $\chi_{00}^{\rm reg}$ is a function which is not
singular at the gap edge; one expects $\chi_{00}^{\rm reg}$ to be
basically equal to the normal state susceptibility, and 
$A\sim(\Delta/E_{F})$. Of course, as the Fermi surface approaches the
van~Hove point, $A$ increases. The dispersion $\epsilon_{p}=-2t(\cos
p_{x} + \cos p_{y})+4t'\cos p_{x}\cos p_{y}-\mu$ with $t'=t/2$ and
$\mu = -1.5t$ has been claimed to describe YBa$_2$Cu$_3$O$_7$
\cite{Si96}. For this dispersion $\chi_{00}^{\rm reg}=0.6/t$ and 
$A = 0.8 (\Delta/t)$. Thus using $t\approx 0.2$eV and $\Delta\sim$
20~meV, we estimate $A\sim 0.1$. The coefficient of the logarithm is quite
sensitive to details of band structure, thus the significance of this
estimate is that it may be reasonable to assume A is small, but not
absurdly so. Note, though, that a key feature of the calculation is
that at $p_0$ and $p_0+Q$ the superconducting gap is at or near its
maximum, so that $A$ is maximal and there is no quasiparticle damping.
Other workers (including \cite{Normand95}) have studied a model with
$t'=0$.  For this model, $p_0$ and $p_0+Q$ are near gap minima, so A
is very small and the damping is large, so the effects we will discuss
below will not be observable.  Thus it may be concluded that within
the present theory the 41 meV peak is very sensitive to details of
band structure and gap anisotropy.

The RPA formula $\chi^{(a,s)} = \chi_{00}/(1-(J\pm
J_{\perp})\chi_{00})$ and Eq. \ref{eq:chinonint} thus imply that the
susceptibility diverges at frequencies $\omega^{a,s}$ given by
\begin{equation}
2\Delta_{p_0}-\omega^{a,s}=2\Delta_{p_0}\exp \left[ -{1 \mp I_\perp\over 
(J\pm J_{\perp})\chi_{max}A}\right]
\label{eq:pole}
\end{equation}
Here $\chi_{max}=\chi_{00}/\left(1-J\chi_{00}\right)$ and $I_\perp =
J_{\perp}\chi_{max}$ have the same meaning as in section 3. One sees
immediately that if the system is not too far from a magnetic
instability the large value of $J\chi_{max}$ may compensate for the
small value of $A$. Also, the argument of the exponential differs for
the symmetric $(1+I_\perp)/(J-J_{\perp})$ and antisymmetric
$(1-I_\perp)/(J+J_{\perp})$ cases. If $J\chi_{max}A<1$ and $I_\perp$ is
not too small, then only the antisymmetric pole is appreciably removed
from the gap edge. Finally, the weight in the resonance, $I^{\pm}=\int
d\omega\chi^{''}$, may be written approximately
\begin{equation}
I^{a,s} = {4\pi(2\Delta_{p_0} - \omega_{a,s})\over J A}
\end{equation}

The weight scales with the difference in energy from the gap edge, and is 
thus much smaller for the symmetric mode.  It is enhanced by the factor 
$1/A \sim 10$. Because precise values for the various quantities are not 
available it is difficult to compare our estimate of the weight precisely 
to experiment.  In Ref. \cite{Liu95} the value $J \sim 80\;{\rm meV}$
was used \cite{Si96}.  The conventions
are those of this paper, so to convert I to the
units used in \cite{Fong96} one must multiply by $2/\pi$.
The result is $I^{a,s} \sim 40\;(1-\omega^{a,s})\;2\Delta_{p_0}$.
It thus seems that weights of order unity or larger
in the antisymmetric channel
are not difficult to obtain.
\section{Conclusions}

We have given expressions which may be useful in interpreting
experiments relating to bilayer coupling in YBa$_2$Cu$_3$O$_{6+x}$. We
have determined the quantum correction to the optic magnon gap in the
insulator, and given the theory of the NMR cross relaxation T$_2$
experiment.  We have shown that a $J_{\perp}\approx 14$~meV is roughly
consistent with all experiments, and that the cross-relaxation
experiment implies that $\chi$ has a very substantial
antiferromagnetic peak, which however is not large enough to have been
observed in recent neutron experiments.

We have shown that a $J_\perp \approx 14\;\;{\rm meV}$ is consistent with
neutron scattering experiments and with an extension to the bilayer
system of the widely - used ``MMP'' analysis
\cite{Millis90,Barzykin95} of NMR. We have also noted that there is at
present no obvious contradiction between neutron scattering and the
MMP scenario for NMR. We have proposed an explanation for the 41 meV
peak observed in the superconducting state of YBa$_2$Cu$_3$O$_7$ and
have argued that in the Y-Ba system at dopings corresponding to
YBa$_2$Cu$_3$O$_{6.63}$ interplane effects will have a significant
effect on the low-energy magnetic dynamics of the underdoped
materials.  Understanding the effects is a crucial problem for future
research.

We thank T. M. Rice, R. Stern and M. Mali for helpful discussions of
NMR, M. Takigawa for drawing our attention to the difference between
NMR and neutron conventions, D. Pines for a critical reading of the
manuscript, B. Keimer for discussions of the neutron experiments,
D. A. Huse for discussion of the $S=1/2$ Heisenberg model, H.~Fukuyama
for discussions of RPA calculations and K. Levin, Q. Si and Y. Zha for
discussions of Ref. \cite{Liu95}.  HM acknowledges the hospitality of
AT\&T Bell Laboratories where this work was begun and AJM and HM thank
the Institute for Theoretical Physics in Santa Barbara where it was
completed.
\newpage

*Address after Sept 1, 1996:  Department of Physics, The Johns Hopkins
University, Baltimore MD 21218.

\begin{figure}[t]

\protect\centerline{\epsfbox{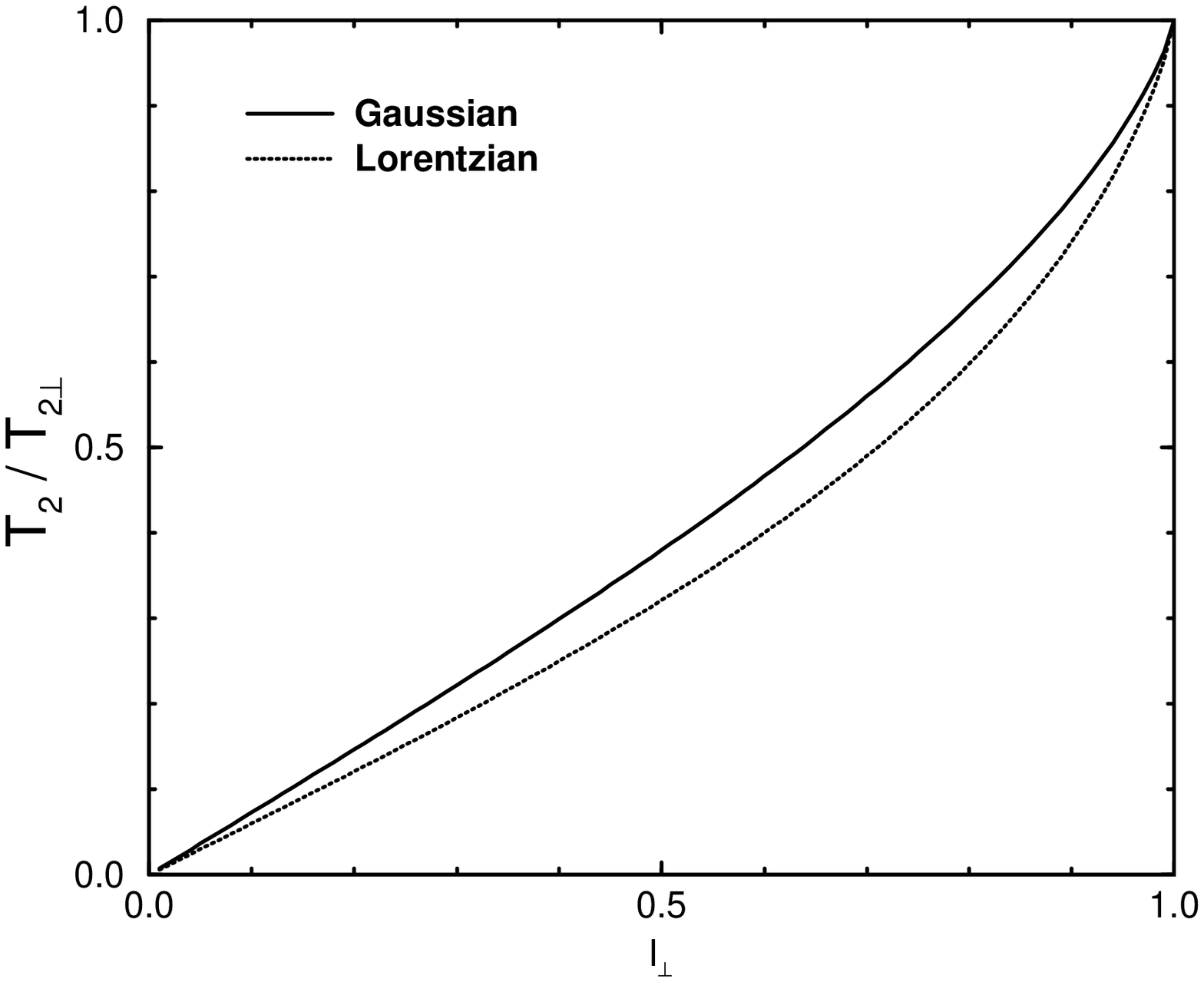}}

\protect\caption{ Ratio of cross-relaxation rate $1/T_{2\perp}$ to 
in-plane 
relaxation rate $1/T_{2}$ plotted versus coupling parameter $I_\perp = 
J_{\perp}\chi_{max}$ for Lorentzian (dotted line) and Gaussian (solid 
line) form factors and calculated from Eqs. (6-9). }

\end{figure}

\begin{figure}[h]

\protect\centerline{\epsfbox{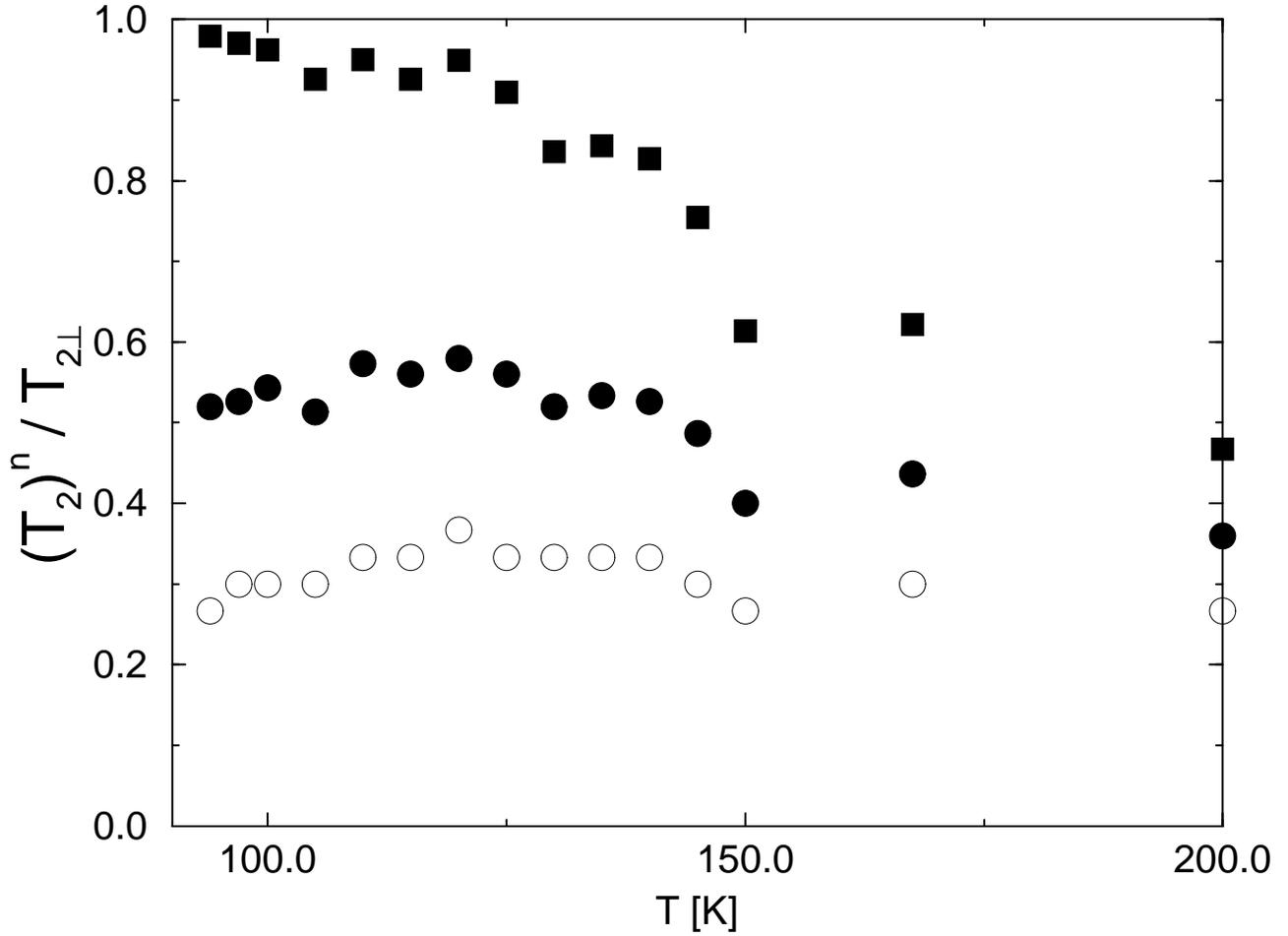}}

\protect\caption{ Experimentally determined ratio of $1/T_{2\perp}$ to 
nth 
power of $1/T_{2}$ for n=1 ($\Box$), 2 ($\bullet$), 3 ($\circ$) in 
arbitrary units. That the n=3 ($\circ$) curve has less temperature 
dependence than the n=2 ($\bullet$) curve suggests
the existence of a growing magnetic correlation length. } 

\end{figure}

\begin{figure}[h]

\protect\centerline{\epsfbox{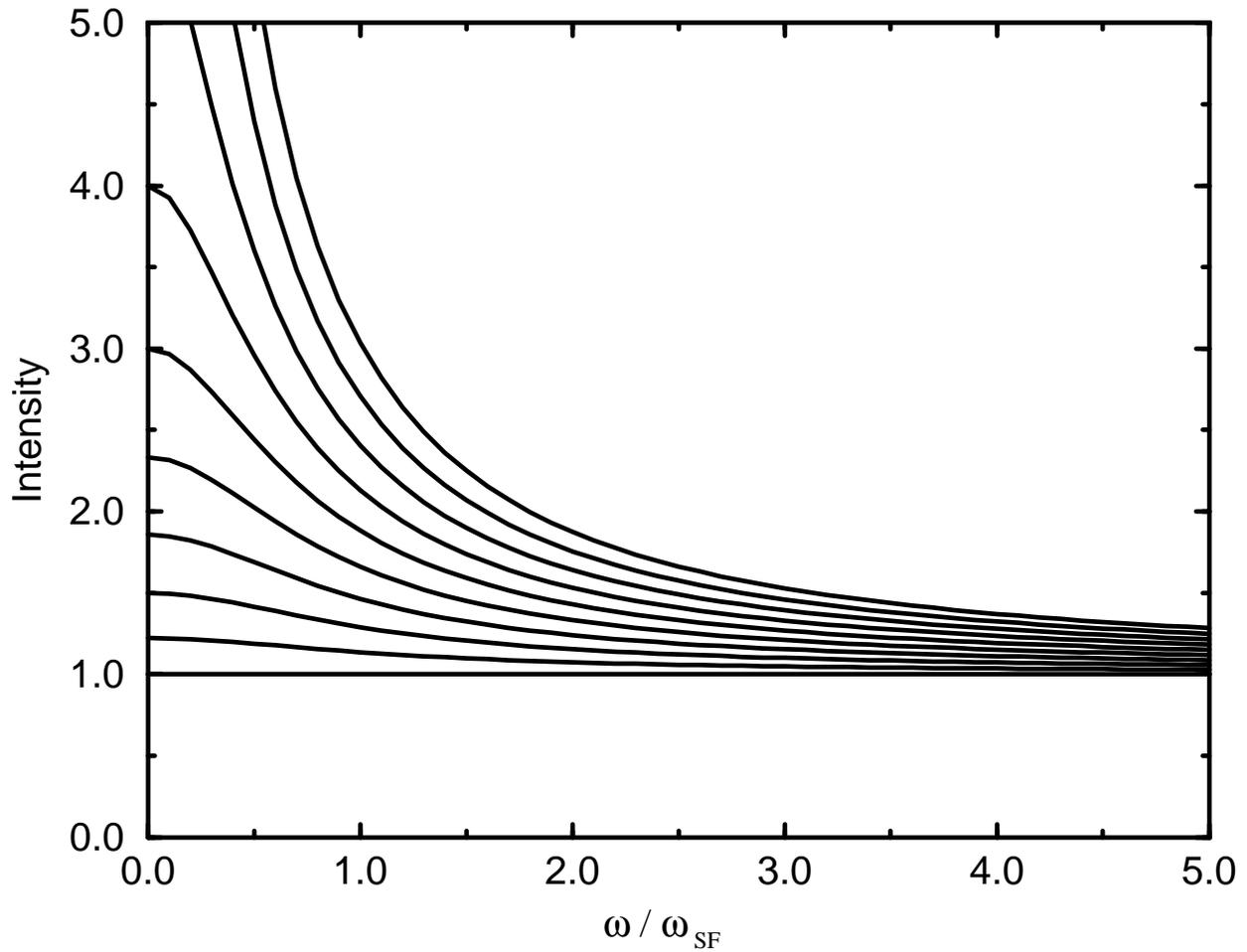}}

\protect\caption{ Calculated ratio of $q$-integrated odd-parity neutron 
absorption to q-integrated even parity neutron absorption, plotted versus 
frequency for $I_\perp = J_{\perp}\chi_{max}$ = 0.0, 0.1 \ldots 0.9. 
$I_\perp=0.0$ corresponds to the lowest curve and $I_\perp=0.9$ to the top 
curve.}

\end{figure}

\end{document}